\documentclass[12pt,a4paper]{article}
\usepackage{latexsym}
\usepackage{amsmath,amscd}
\usepackage{amsfonts,amssymb}
\input{psfig.tex}

\newcommand*{\npar}{\bigskip\par\noindent}

\newcommand*{\trm}[1]{m_{\perp #1}}
\newcommand*{\JP}{\ensuremath{J/\psi}}
\newcommand*{\ac}{\ensuremath{\bar c}}
\newcommand*{\qq}{\ensuremath{q\bar q}}
\newcommand*{\cc}{\ensuremath{c\bar c}}
\newcommand*{\zh}{\ensuremath{\hat{\zeta}}}

\newcommand*{\beq}{\large\begin{equation}}
\newcommand*{\eeq}{\end{equation}\normalsize}
\newcommand{\nsz}{\normalsize}
\newcommand*{\bmath}{\large $$}
\newcommand*{\emath}{$$ \normalsize}

\begin{document}

%
%
\vspace*{1.0cm}
\leftline{\large{DESY 98--174}}
\leftline{\large{November 1998}}

\vspace*{1.0cm}
\centerline{\Large{\bf The Distribution of Constituent Charm Quarks}}
\centerline{\Large{\bf in the Hadron}}

\vspace*{1.0cm}
\centerline{\large Yu.A.Golubkov}

\bigskip
\centerline{{\em Moscow State University, Moscow, Russia}}
\centerline{{\em and}}
\centerline{{\em Deutsches Elektronen -- Synchrotron DESY, Hamburg}}

\bigskip
\centerline{e-mail:~golubkov@npi.msu.su,~golubkov@vxdesy.desy.de}

\begin{abstract}
Using a statistical approach in the framework of non--covariant
perturbation theory
the distributions for light and charmed quarks in the hadron
have been derived, taking into account the mass of the charmed quark.
The parameters of the model have been extracted from
the comparison with NA3 data on hadroproduction of \JP ~particles.
A reanalysis of the {\em EMC} data on charm
production in muon--nucleon scattering has been performed.
It has been found in comparison
with the conventional source of charmed quarks from photon-gluon fusion,
that the EMC data indicate the presence of an additional contribution
from deep-inelastic scattering on charmed quarks at large $x$.
The resulting admixture of the Fock states, containing charmed quarks
in the decomposition of the proton wave function is
of the order of 1\%. The approach presented
for the excitation of the Fock states with charmed quarks
can also be applied to states with beauty quarks as well as
to the hadronic component of the virtual photon (resolved photon component).
\end{abstract}

\bigskip
\section{Introduction}
\label{introd}

The production of heavy flavours in lepton--hadron and hadron--hadron
collisions is a very important tool for a quantitative test of
QCD and
for searches for new physics.
Due to the presence of the point--like
probe particle (lepton) and the possibility to control the QCD scale
of the hard sub--process
deep inelastic scattering has a number of advantages
in comparison with hadronic reactions in the analysis
of charm production. The QCD--based parton model has been
remarkably successful in describing a wide
variety of high--energy processes involving energy scales much
larger than the masses of the known particles and the partons
themselves.
Many analysis of charm production performed within
the framework of the parton model assume that the hadron consists
only of massless or approximately massless
partons (gluons, $u$, $d$ and $s$ quarks).
The heavy quarks (charm, bottom) are treated as massive objects
which are external to the hadron.
In the DIS--neutral current reactions this kind of consideration
leads naturally to the $O(\alpha_s^1)$ "photon--gluon fusion"
(PGF) mechanism $\gamma g\,\rightarrow\,c\bar c$,
as the main mechanism for heavy quark production.
In hadronic collisions the analogue "parton--parton fusion" processes,
$gg\,\rightarrow\,c\bar c,\ q\bar q\,\rightarrow\,c\bar c$,
are expected to contribute.
These parton fusion processes are flavour creating (FC),
since the heavy flavour is created by the interaction with
a light constituent of the hadron.

\bigskip
The existing experimental
data on $\mu p$ collisions \cite{EMC82} show some irregularities
which are inconsistent with the PGF
predictions.
The experimental observation of a deviation,
of the charm distribution at large pseudorapidities
(i.e. charm production close to the direction of the proton
beam) from the
conventional predictions in $ep$ scattering have also been
reported by ZEUS experiment at HERA collider \cite{Vancouver98}.
In hadronic collisions the interpretation of the data
on open charm production
within the standard parton--parton fusion scheme followed by
hadronization of charmed quarks meets problems
for charmed particles at large $x_F$.
In this region the charm distributions are harder than then
predictions of the factorization approach. Furthermore, the yield of
charmed particles containing the valence quarks of the projectile
significantly exceeds the yield of their anti--particles.
Models considering recombination
of the newly created charmed quark with one of the valence quarks
of the projectile \cite{Liho81,Golub82} or string fragmentation
\cite{Pythia57} can improve the situation with open charm production.
In both approaches a part of the proton remnant
momentum is imparted to the final state charmed particle increasing its
momentum and improving the agreement with the experimental
observations.
However these models have problems with
describing $\JP$ and double--$\JP$ production as well as the
$A$--dependence of the charm production cross section in
hadron--nucleus collisions at large $x_F$
\cite{Boreskov93,Badier83,Vogt91}.

\bigskip
A part of the discrepancies between data and models can  be
resolved by introducing
the scheme of {\em flavour excitation} (FE),
which assumes that also heavy quarks can be constituents
of the hadron.
We note that considering heavy quarks as external to the hadron is
appropriate when the characteristic scale of the process ($\mu$) is
less or of the order of the mass of the heavy quark, i.e.,
$\mu\,\lesssim\,m_Q$. This condition holds for $c$ and $b$ quarks
for most fixed--target experiments. The $ep$
collider HERA gives an opportunity to investigate the heavy quark
production at $Q^2$ scales much larger than $4\,m_Q^2$.
At such scales it seems justified
to consider charm (and bottom) quarks as light
objects.

\bigskip
The so called {\em "variable flavour number scheme"},
suggested in papers \cite{Olness88,Aivazis94,Olness95}, combines
both, the {\em FC} and {\em FE} schemes and ensures a "soft"
transition between the two production mechanisms.
A more sophisticated approach for the FC mechanism
re-sums effectively the large logarithms of the
type $\left [\alpha_s(\mu)\ln(\mu^2/m_Q^2)\right ]^n$ which limit
the validity of the conventional FC mechanism to the region
$\mu\,\sim\,O(m_Q)$.
It has been shown in these papers that the contribution from the
scattering on a constituent charm quark of the proton (process of order
$O(\alpha_s^0)$) becomes more important than PGF at
$x\,\gtrsim\,0.1$ already at surprisingly low $Q^2\,\gtrsim\,20-30\
GeV^2$.
The authors used splitting functions and standard
~distributions for massless partons, including
heavy flavours and introduced the "slow--rescaling" variable
$x\ \rightarrow\,x\,\left [1+(m_Q/Q)^2\right ]$ to account for the
charmed quark mass.
However perturbative QCD requires the scale $\mu$ to be large
and needs as input the initial parton distributions
for their evolution.
The initial distributions for heavy quarks are not necessarily similar
to the distributions of light partons due to non--zero quark mass,
which is comparable with the QCD scale. This kind of consideration
is closely related to the old question: which type of high order
QCD corrections have to be assigned to the matrix element and
which due to QCD evolution of the parton distributions. So far we have
no clear understanding of this problem.

\bigskip
The authors of \cite{Brodsky80} have suggested a procedure to obtain
the distribution of {\em massive} charmed quarks in the hadron.
They considered the proton wave function decomposition which may
contain the \hbox{$\vert\,uud\ac c>$} Fock state component
called "intrinsic charm" (IC).
Such a state may appear as a quantum fluctuation
of the hadron wave function and may become free
in interactions with substantial momentum transfer.
In this case the proton is described as a decomposition in terms of
colour--singlet eigenstates of the free Hamiltonian
$\vert uud>$, $\vert uudg>$, $\vert uudq\bar q>$, $\ldots$.
Over a sufficiently short time the proton can contain Fock states
of arbitrary complexity, including pairs of charmed quarks.
In the proton rest frame the life--time  $\tau$ of such fluctuations
is of the order of the nuclear time $\sim\,R_h$, where $R_h$ is the
hadronic size.
On average there are in addition to the valence quarks
extra partons (gluons and \qq ~pairs).
In the infinite momentum frame a partonic fluctuation
will be "frozen" and  can be observed,
e.g., in lepton--hadron scattering.
The charmed quarks are heavy
objects and their life--time is much smaller than for light partons.
Thus, on average,  the admixture of heavy quark pairs is expected to
be small, $\sim\,(m_q/m_Q)^2$.
Because the quantum fluctuations in the initial proton
is defined by the colour field self--interaction,
the structure of Fock states of the proton
can be considered independently of the hard interaction,
providing the initial non--perturbative parton distributions.
These distributions will evolve in hadron-- or lepton--hadron
collisions due to large momentum transfer.

\bigskip
In the present paper we modify and generalize the statistical approach
to the Fock state hadron structure with heavy quarks,
suggested in Ref.\cite{Brodsky80,Kuti71,Golub96},
within the framework of the non--covariant perturbation theory.
We obtain scaling expressions for the heavy and light parton
distributions in the infinite momentum frame.
We calculate the charmed structure function of the proton,
$F_2^{(c)}(x,Q^2)$, taking into account the QCD radiative corrections
order of $\alpha_s^{(1)}$ as well as the mass corrections
caused by the non--zero values of the $c$ quark and the proton masses.
We use the experimental data for $\pi A\,\rightarrow\,\JP X$ \cite{Badier83}
and $\mu p\,\rightarrow\,\mu c\bar c X$ \cite{EMC82}
to evaluate the parameters of the model and present the relative
contributions of the {\em PGF} and the {\em IC} mechanisms to the
charm structure function of the proton. Note that we use the terms
{\em FE} and {\em intrinsic charm (IC)} for the
charm production mechanisms involving constituent charmed quarks
of the proton.

\section{Description of the model} 
\label{model}

\subsection{General features}
\label{arbin}

In QCD high energy hadrons are coherent superpositions (Fock state
vectors) of quarks and gluons. Note that the lifetime of the
fluctuation
$\Delta t\,\sim\,1/\Delta E\ \approx\ 2\,P_h/\left (M^2-m^2\right )$
($P_h$ is the momentum of hadron, $m$ is its mass and $M$ is the
mass of the fluctuation) can be very large at high energies even
for large mass values of the fluctuation.

Based on above picture of the proton and using the statistical
approach, the authors of Ref.\cite{Kuti71} achieved a good
description of the proton structure function.
In Ref.\cite{Golub96} we have presented a statistical
consideration of the hadron structure and obtained non--invariant,
i.e. frame dependent, expressions for parton distributions.
In principle the frame dependence can take place at sufficiently
low energies while in the infinite momentum frame
one expects invariant expressions.

We take all partons on the mass shell
and use non--covariant perturbation theory.
Thus we consider a hadron as a statistical system which
consists of $N$ quarks carrying quantum numbers of a hadron,
two charmed quarks $c$, $\bar c$ and a system of $n$ light partons
(gluons and quarks) carrying in total the quantum number of the
vacuum.

In non--covariant perturbation theory
the probability to produce a $m$ particle final state
in the case of the instant interaction potential looks as follows
\cite{Landau3}:

\beq
\label{perprob1}
dW^{(m)}\ \sim\ \frac{\vert H_{int}\vert ^2}{\left (E_{fin}\,-
\,E_h\right )^2}\,
\delta\left (\vec P_{fin}\,-\,\vec P_h\right )\,d\Phi_{fin}^{(m)},
\end{equation}\nsz

\noindent where $d\Phi^{(m)}_{fin}$ is the element of the $m$ particle
phase space, $m\,=\,N+2+n$, $\vec P_h$ and $E_h$ are the momentum
and energy of the considered hadron; $\vec P_{fin}$ and $E_{fin}$ are
the momentum and energy of the final state partonic fluctuation,
respectively; $d\Phi^{(m)}_{fin}$ describes the Lorentz invariant
phase space:

\beq
d\Phi_{fin}^{(m)} \ = \ \prod_{i=1}^m\frac{d^3p_i}{\varepsilon_i}\,.
\end{equation}\nsz

The $\delta$--function ensures conservation of the total
3--momentum.

Due to the sharp cut--off for the transverse momenta of partons
it is sufficient to consider only the longitudinal phase space:

\beq
\label{phspa2}
\frac{d^3p}{\varepsilon}\ \rightarrow\ \frac{d\xi}{\sqrt{\xi^2+\mu^2}}\,,
\end{equation}\nsz

\noindent where $\xi\,=\,p_z/P_h$, $\mu\,=\,\trm{}/P_h$
and $\trm{}$ is the transverse mass of the parton.

Following the parton model we assume
independent primordial distribution of each parton and replace:

\beq
\label{intham}
\vert H_{int}\vert ^2\,d\Phi^{(m)}_{fin}\ \rightarrow
\ \prod_{i=1}^m \,\rho_i(\xi_i)\,d\xi_i\,,
\end{equation}\nsz

\noindent where $\rho(\xi)$ is the probability density
to observe the parton with the momentum fraction $\xi$.
Therefore, the probability to observe an $m$ parton Fock state
looks as:

\beq
\label{probhat}
W^{(m)}\ =\ \int_0^1\,\prod_{i=1}^m\,d\xi_i \,\rho_i(\xi_i)
\,\delta\left (1-\sum_{j=1}^m\,\xi_j\right )\,.
\end{equation}\nsz

The omitted common factors will be incorporated in the general
normalization.
The integration over $m$ parton momenta
can be performed \cite{Kuti71} with help of the integral representation
of the $\delta$--function:

\beq
\label{dltexpan}
2\,\pi\,\delta(x)\ = \ \int_{-\infty}^{+\infty}\,d\nu\,e^{i\nu x}\,.
\end{equation}\nsz

\bigskip
If we perform the integration over all $\xi_i$, we obtain:

\beq
\label{furhat}
W^{(m)}\ =\ \frac{1}{2\pi}\,\int_{-\infty}^{+\infty}\,d\nu
\,e^{i\nu}\,\prod_{i=1}^{m} \,\rho_i(\nu)\,.
\end{equation}\nsz

\noindent Here we extended integration of $\xi$ to $\infty$
due to the presence of the fast oscillating exponent and defined
as $\rho_i(\nu)$ the Fourier transformation
of the parton density $\rho_i(\xi)$:

\bmath
\rho(\nu) \ = \ \int_0^{\infty}\,d\xi\,\rho_i(\xi)\,e^{-i\nu\xi}\,.
\emath

Following \cite{Kuti71} we introduce different probability
densities for valence quarks, charmed quarks, gluons and light sea
quarks, $\rho_v$, $\rho_c$, $\rho_g$ and $\rho_q$, respectively.
Because all sea light partons of the same type have
the same distributions we have to sum over
all possible permutations of the $n$ light sea partons (gluons
and \qq ~pairs separately). Thus, we have for the probability
to observe a Fock state with $N$ valence quarks, one \cc ~pair,
$n_g$ gluons and $n_q$ pairs of light sea quarks
($n\,=\,n_g\,+\,2n_q$) the expression:

\beq
\label{probfock}
W_N^{(n)}\ =\ \frac{1}{2\pi}\,\int_{-\infty}^{+\infty}\,d\nu\,e^{i\nu}\,
\left [\rho_v(\nu)\right ]^N\,\rho_{\cc}(\nu)
\,\sum_{n_g+2n_q=n} \,\frac{\rho_g(\nu)^{n_g}}{n_g!}
\,\frac{\rho_q(\nu)^{2n_q}}{(2n_q)!}\, ,
\end{equation}\nsz

\noindent where $\rho_{c\bar c}$ is the probability to create
$c\bar c$ pair.

The factors $1/n_g!$ and $1/(2n_q)!$ take into account
the non\-dusting\-uish\-ability of gluons and quarks,
respectively.
To obtain the total probability for all Fock states,
containing the \cc ~pair one needs to perform a summation over
$0<n<\infty$. This summation can be carried out by using
the properties of binomial sums:

\beq
\label{sumqg}
\sum_{n_g+2n_q=n}\,\frac{\rho_g^{n_g}}{n_g!}
\,\frac{\rho_q^{2n_q}}{(2n_q)!}
\ =\ \frac{1}{2}\,\left [\frac{\rho_+^n\,}{n!}
+\,\frac{\rho_-^n}{n!}\right ];
\ \ \ \rho_{\pm}\,=\,\rho_g\,\pm\,\rho_q\,.
\end{equation}\nsz

The general form of the statistical sum is \cite{Takasugi79}:

\begin{eqnarray}
\label{genorm}
Z_N^{(c)} & = & \ \sum_{n=0}^{\infty}\,W_N^{(n)}\\
 & = & \frac{1}{2\pi}\,\int_{-\infty}^{+\infty}\,d\nu\,e^{i\nu}
\ \left [\rho_{v}(\nu)\right ]^N\ \rho_{c\bar c}(\nu)
\ \exp \left [\rho_g(\nu)\right ]
\ \cosh \left [\rho_q(\nu)\right ].\nonumber
\end{eqnarray}

It is clear from Eq.(\ref{probhat}), that the distribution
$P_i(\xi)$ for the $i$th parton can be obtained if one omits
the integration over momentum of the considered parton.
In general, one-- or many--particle distributions can be derived
from the statistical sum taking the functional derivative
of the desired function(s) \cite{Takasugi79}.
Thus, the inclusive distribution
of light partons and the distribution of the \cc ~pair, normalized
to unity, looks as follows:

\beq
\label{gendist}
\begin{array}{lllll}
P_i(\xi) & = & \frac{1}{Z_N^{(c)}}\,\rho_i(\xi)\
\frac{\delta\,Z_N^{(c)}}{\delta\,\rho_i}
& \equiv & \frac{1}{Z_N^{(c)}}\,\rho_i(\xi)\ C_i(1-\xi)\,,\\
\\
P_{\cc}(\xi_c,\xi_{\bar c}) & = & \frac{1}{Z_N^{(c)}}
\,\rho_{\cc}(\xi_c,\xi_{\bar c})\
\frac{\delta\,Z_N^{(c)}}{\delta\,\rho_{\cc}}
& \equiv & \frac{1}{Z_N^{(c)}}\ \rho_{\cc}(\xi_c,\xi_{\bar c})
\ C_{\cc}(1-\xi_c-\xi_{\bar c})\,,
\end{array}
\end{equation}\nsz

\noindent where, $C_i(1-\xi)$ and $C_{\cc}(1-\xi_c-\xi_{\bar c})$
are the correlation functions, ensuring the momentum conservation.

\subsection{Probability densities and parton distributions.}

The origin of the \cc ~pair is the same as of light sea quark pairs,
namely the splitting of the gluon into a virtual \cc ~pair,
$g\,\rightarrow\,\cc$.
In principle, at sufficiently small energies,
when $\xi_c\,P_h\ \le\ \trm{c}$,
one can expect $\rho_c\,\approx\,const$,
but in the limit $P_h\,\rightarrow\,\infty$, one has case
$\xi\gg\mu_{c}$ and therefore a situation
similar to light sea quarks.
We shall compare the model with fixed target experimental data,
neglecting the transverse mass of the charmed quarks
for the projectile hadron.
Thus, following \cite{Kuti71} we can represent
the probability densities $\rho(\xi)$ as:

\beq
\label{rhoval}
\begin{array}{lll}
\rho_v(\xi) & \propto & \frac{\xi^{\alpha}}{\sqrt{\xi^2+\mu_v^2}}
 \ \approx \ \xi^{\alpha-1}\, ,\\
\\
\rho_g(\xi) &    =     & \frac{a_g}{\sqrt{\xi^2+\mu_g^2}}\, ,\\
\\
\rho_q(\xi) &    =     & \frac{a_q}{\sqrt{\xi^2+\mu_q^2}}\, ,\\
\\
\rho_{c}(\xi) & \propto  & \frac{1}{\sqrt{\xi^2+\mu_c^2}}
 \ \approx \ \frac{1}{\xi}\, ,
\end{array}
\end{equation}\nsz

\noindent for valence quarks, gluons, light sea quarks
and charmed quarks, respectively,
with $a_g$ and $a_q$ being unknown constants.

In the infinite momentum frame we neglected the transverse
mass in the probability densities
of the valence and charmed quarks, Eq.(\ref{rhoval}).
At the same time for sea partons we hold temporarily the term $\mu$
in the denominators to perform later the Fourier transformations.
In the final expressions we shall put $\mu\,\rightarrow\,0$.

From experiment we know that
the valence quark momentum distribution at small $\xi$
is approximately proportional to $1/\sqrt{\xi}$, i.e.,
$\alpha\,=\,0.5$. We shall use this value in our comparison
with experimental data, but in the formulae we use the general expression
(\ref{rhoval}).

Let us consider the energy denominator in Eq.(\ref{perprob1}).
Taking into account the momentum conservation
{\hbox{$P_h\,=\,\sum_i\,p_i$}},
from the light--cone expansion in the infinite momentum frame
one obtains:

\beq
\label{lcexpan}
E_{fin}\,-\,E_h\ \approx\ \frac{1}{P_h}
\,\left (M_h^2\,-\,\sum_{i=1}^m\frac{\trm{i}^2}{\xi_i}\right )\,.
\end{equation}\nsz

Because the charmed quark transverse mass is much larger
than $M_h$ (the mass of the hadron) and than the transverse mass
of the light partons $\trm{i}$,
the energy denominator will be proportional to \cite{Brodsky80}:

\beq
\label{cqfact}
\frac{1}{\left (E_{fin}\,-\,E_{in}\right )^2}\ \propto \ \frac{\xi_c^2
\xi_{\bar c}^2} {(\xi_c\,+\,\xi_{\bar c})^2}\,.
\end{equation}\nsz

The validity of the approximation for the light--cone expansion
of the energy denominator for Fock states with heavy quarks
has been considered in Ref.\cite{Golub96}.

We can introduce the expression (\ref{cqfact}) in the definition
of the probability density $\rho_{\cc}$ to observe the pair \cc ,

\bmath
\rho_{c\bar c}(\xi_c,\xi_{\bar c})\ \equiv\ \frac{\xi_c^2
\xi_{\bar c}^2} {(\xi_c\,+\,\xi_{\bar c})^2}
\,\rho_c(\xi_c)\,\rho_{\bar c}(\xi_{\bar c})
\ =\ \frac{\xi_c\xi_{\bar c}} {(\xi_c\,+\,\xi_{\bar c})^2}\, .
\emath

Similar to case of the valence quarks
we keep the more general form for \cc ~probability density
in further formulae:

\beq
\label{cprob}
\rho_{c\bar c}(\xi_c,\xi_{\bar c})\ =
\ \frac{\xi_c^{\beta}\,\xi_{\bar c}^{\beta}}
{(\xi_c\,+\,\xi_{\bar c})^2}\, .
\end{equation}\nsz

We introduced in above formula the phenomenological parameter $\beta$
to take into account possible deviation of the charm distribution
from the phase space approximation Eq.(\ref{rhoval}) at moderate energies
or momentum transfers. In analytical expressions we use generalized formula
Eq.(\ref{cprob}). But in numerical calculations and in comparison with data
in present paper we use the phase space approximation and hold $\beta\,=\,1$.

Therefore from Eq.(\ref{genorm}) we obtain for the statistical sums
for Fock states with ($Z_N^{(c)}$) and without ($Z_N$) the \cc ~pair
in the limit $\mu\,\rightarrow\,0$:

\bigskip\vbox{
\beq
\begin{array}{lll}
Z_N^{(c)} & = & \frac{1}{2\pi}
\,\int_{-\infty}^{+\infty}\,d\nu\,e^{i\nu}
\ \left [\frac{\Gamma (\alpha)}{\nu^{\alpha}}\right ]^N
\ \frac{1}{\nu^g}\ \frac{\Gamma(2\beta)\,f_c(\beta)}{\nu^{2\beta}}\,, \\
\\
Z_N\ & = & \frac{1}{2\pi}
\,\int_{-\infty}^{+\infty}\,d\nu\,e^{i\nu}
\ \left [\frac{\Gamma (\alpha)}{\nu^{\alpha}}\right ]^N
\ \frac{1}{\nu^g}\, ,
\end{array}
\end{equation}\nsz
}

\noindent where $\Gamma(x)$ is the gamma function and $g\,=\,a_g+a_q$
is the unknown parameter of the model, characterizing the level of
the sea in the hadron considered.

Note here, that the integral for light sea partons is
$\sim\,1/\mu^{g}$ and logarithmically diverges in the limit
$\mu\,\rightarrow\,0$.
But this divergence can be incorporated in the general normalization,
as seen from Eq.(\ref{gendist}) and is not important.

The analytical expressions for the statistical sums
with and without the \cc ~pair are:

\large
\vbox{
\begin{eqnarray}
\label{ccnorm}
Z_N^{(c)} & = & \frac{\left [\Gamma (\alpha)\right ]^N
\,\Gamma(2\beta)\,f_c(\beta)}
{\Gamma \left (\alpha N+2\beta+g\right )}\nonumber\\
\nonumber\\
Z_N\ & = & \frac{\left [\Gamma (\alpha)\right ]^N}
{\Gamma\left (\alpha N+g\right )}\\
\nonumber\\
f_c(\beta) & = & \int_0^{\pi/2}\,d\varphi\,
\frac{\left ( \sin\,\varphi\,\cos\,\varphi\right )^{\beta}}
{\left (\sin\,\varphi\,+\,\cos\,\varphi\right )^{2\beta+2}}\,.\nonumber
\end{eqnarray}
}
\nsz

For integer values of $\beta$ the function $f_c(\beta)$ is:
$f_c(0)\,=\,1$, $f_c(1)\,=\,1/6$, $f_c(2)\,=\,1/30$
and $f_c(3)\,=\,1/140$. For arbitrary
values of $\beta$ the integration can be performed numerically.

\bigskip
The result for the parton momentum distributions for the valence quarks,
sea partons and charmed quarks is:

\beq
\label{vcrf}
\begin{array}{lll}
V(\xi) & = & \frac{Z_{N-1}^{(c)}} {Z_N^{(c)}}\ \xi^{\alpha-1}
\ (1-\xi)^{\alpha (N-1)+2\beta +g-1}\,,\\
\\
S(\xi) & = & g\ \xi^{-1}\,(1-\xi)^{\alpha N+2\beta+g-1}\,,\\
\\
P_{\cc}(\xi_c;\xi_{\ac}) & = &
\frac{\xi_c^{\beta}\xi_{\ac}^{\beta}}{\left (\xi_c+\xi_{\ac}\right )^2}
\,(1-\xi_c-\xi_{\ac})^{\alpha N+g-1}\,,\\
\\
c(\xi) & = & \frac{Z_N}{Z_N^{(c)}}\ \xi^{\beta}
\ (1-\xi)^{\alpha N+\beta+g}\ J_N^{(c)}(\xi)\,,\\
\\
J_N^{(c)}(\xi) & = & \int_0^1\,dy
\,\frac{y^{\beta}\,(1-y)^{\alpha N+g-1}}
{\left [ \xi\,+\,(1-\xi)y\right ]^2}\,.
\end{array}
\end{equation}\nsz

The valence and $c$ quark distributions are normalized to unity
$\int_0^1\,d\xi\,P(\xi)\,=\,1$.

Despite the fact that the mass of charmed quarks does not enter
directly the final expression for the probability of the Fock
state, we see from Eq.(\ref{vcrf}) that due to the factor
$\xi^{\beta}$ with $\beta\,>\,0$ the distribution of charmed
quark is much harder than a distribution of light sea parton.
The origin of this hardness is the large value
of the charmed quark mass. This effect must be taken into account by
any phenomenological parameterization of the initial (not QCD evolved)
charmed quark distribution.

If the $SU(3)$ symmetry in the sea is broken
one can introduce a suppression factor $\lambda_s$ for strange
quarks and obtain:

\begin{eqnarray}
\label{seadist}
S_u(\xi)\,=\,S_{\bar u}(\xi)\,=\,S_d(\xi)\,=\,S_{\bar d}(\xi) & = &
\frac{1}{4\,+\,2\lambda_s}\,S(\xi)\,, \nonumber \\
 \\
S_s(\xi)\,=\,S_{\bar s}(\xi) & = &
\frac{\lambda_s}{4\,+\,2\lambda_s}\,S(\xi)\,. \nonumber
\end{eqnarray}

Since the total probability for the excitation of a \cc~pair
in the hadron is unknown we assumed the normalization to be unity.
The final result can be obtained by multiplication with the
factor $N_c$, which can be extracted from the experimental data.
In the next sections the model is compared with the experimental
data in order to evaluate the free parameters $N_c$ and $g$.

Let us make a comment about the excitation of the states with beauty quarks.
If we consider $b$ quarks within the framework of this approach, we obtain
again the expressions (\ref{cprob}) and (\ref{vcrf})
neglecting the mass of the charmed quark in comparison with
the much larger mass of the beauty quark.
In this case, the charmed quarks have to be treated as massless
partons of the sea.

One can apply this model also to the excitation of the heavy flavours
in the virtual photon ("resolved photon") in $e^+e^-$ annihilation or
in photoproduction. To obtain the necessary distributions
it is enough to omit the term $\left [\rho_{v}(\nu)\right ]^N$
in Eq.(\ref{genorm}) because, obviously, the photon does not contain
valence quarks.

\section{Comparison of the model with \JP ~hadroproduction}
\label{jpsfit}

It was noted in the introduction, that there
exist a number of hadronization
mechanisms which more or less successfully describe
the inclusive  open charm yield at large $x_F$.
These schemes incorporate some aspects of the hadronization process,
but there is no commonly accepted mechanism based on a well
founded theoretical approach, which describes all available data.
The uncertainties in the existing
hadronization models are too large to perform a direct evaluation
of the parameters of the model for charm quark distribution
presented here from the experimental data on open charm production.

On the other hand the experimental data from NA3 \cite{Badier83}
on the production of \JP ~particles in hadron--platinum collisions
indicate an unusual production mechanism.
The authors of NA3 identified two different components in the $x_F$
distribution of \JP~mesons: a "hard" component with the usual
A--dependence $\sigma_A\,\sim\,A\,\sigma_N$ and a "diffractive"
component with a
weaker A--dependence, namely, $\sigma_A\,\sim\,A^{0.77}\sigma_N$
for incident pions and
$\sigma_A\,\sim\,A^{0.71}\sigma_N$ for incident protons.
The relative contribution
of the "diffractive" component is $\approx\,0.20$ and $\approx\,0.30$
for pion and proton projectiles, respectively. The "hard" component,
as shown in the same paper \cite{Badier83}, can be
described well by the conventional QCD parton-parton fusion mechanism.
The linear A--dependence agrees well with predictions
of the QCD factorization theorem valid for hard processes.

The unusual A--dependence of the "diffractive" component can naturally
be described in the model of intrinsic charm \cite{Brodsky80} and the
model \cite{Boreskov93} based on the Gribov approach to particle
interactions with nuclei. It can be shown \cite{Boreskov93}
that for \JP ~production at NA3 energies,
$P_{lab}\,\approx\,(150\,-\,300)\ GeV$,
a deviation from the $A^1$ behaviour for inclusive spectra
only for those components of the initial partonic configuration,
which contain the heavy state.
Therefore, we can ascribe the "diffractive" component in \JP ~distribution,
seen in NA3 experiment, to the interaction of the hadronic Fock state,
containing a \cc ~pair and to use these data to evaluate the parameters
of our model.

To estimate the longitudinal distribution of \JP ~particles we use
the recombination model \cite{Hwa77}.
In this model the differential cross section for \JP ~production
can be written as:

\beq
\label{sigrec}
\frac{d\sigma}{dx_F} \ =\ \sigma_{tot}\,\int\,d\xi_1\,d\xi_2
\,S_{\cc}(\xi_c,\xi_{\ac})\,R(\xi_c,\xi_{\ac};x_F)\,,
\end{equation}\nsz

\npar
where, $S_{\cc}(\xi_c,\xi_{\ac})$ is the two--particle distribution
of $c$ and $\ac$ quarks with momentum fractions $x_c$ and $x_{\ac}$,
respectively; $R(\xi_c,\xi_{\ac};x_F)$ is the recombination function
describing the probability for two quarks to coalesce in the final
\JP ~meson with momentum fraction $x_F$. In the simplest case

\bmath
R(\xi_c,\xi_{\ac};x_F)\ =\ \delta(x_F\,-\,\xi_c\,-\,\xi_{\ac}),
\emath

\noindent ensuring longitudinal momentum conservation.
In principle, for charmed particles, the primordial transverse momenta
of the initial $c$ quarks can reach large values ($\geq \, 1\ GeV$)
and have to be taken into account (see, e.g., \cite{Golub82}).
For the aim of the present paper this is not important. Since we consider
only longitudinal distributions the integration over transverse momenta
is included in the total normalization on experimental data.
As a result, from Eqs.(\ref{vcrf}), (\ref{sigrec})
one obtains for the $x_F$ distribution of $\JP$ particles
the following expression:

\beq
\label{jpsec}
\frac{d\sigma(J/\psi)}{dx_F} \ = \ \sigma_{tot}^{exp}
\ \frac{Z_N}{Z_N^{(c)}}\ \frac{\left [\Gamma(\beta+1)\right ]}
{\Gamma(2\beta+2)}\ x_F^{2\beta-1}\ (1-x_F)^{\alpha N+g-1}\,.
\end{equation}\nsz

\begin{figure}[htb] 
\centerline{\hbox{
\psfig{figure=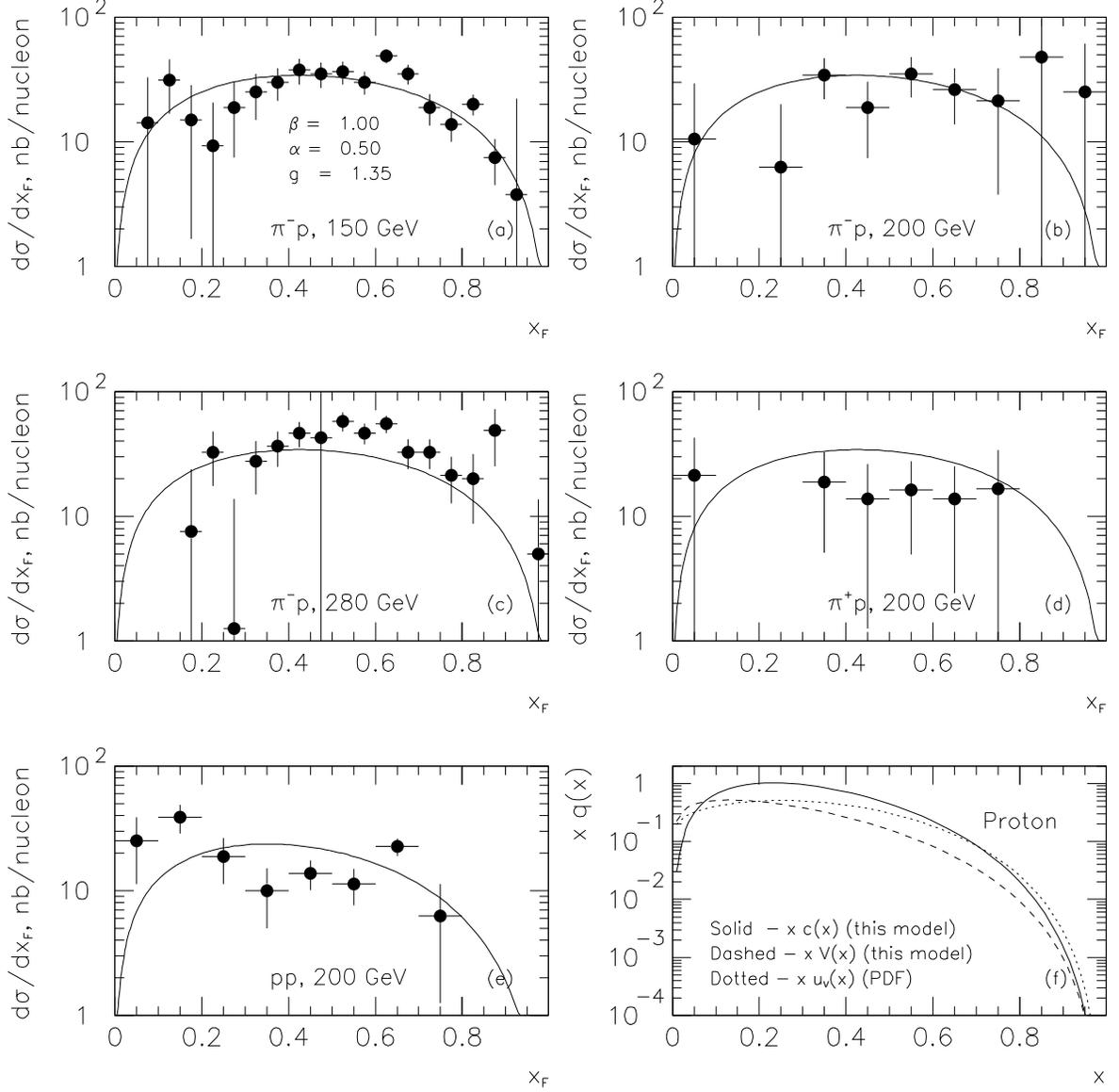,bbllx=0.7cm,bblly=5.2cm,%
bburx=19.0cm,bbury=23.0cm,clip=t,height=16.0cm}
}}
\caption{\label{psifit_1}
Results of a fit to the NA3  data
for $\JP$ production in $\pi\,p$ and $p\,p$ interactions (a-e)
using the model described in the text.
The fit has been performed for $\pi N$ collisions only.
(f) The $x\,q(x)$ distributions in the proton of the $c$ (solid),
valence $u$ (dashed) quarks in the model
and the {\em PDF} parameterization for valence $u$ quark (dotted).
}
\end{figure}

For the fit we used only $\pi\,N$ data from \cite{Badier83}
assuming for the number of valence quarks $N\,=\,2$. We do not use the
$pp\,\rightarrow\,\JP$ data because we actually don't know the
probability for the \cc ~pair to form a \JP ~meson.
This probability can be different for incoming $\pi$--mesons and
protons.
Therefore, expression (\ref{jpsec}) has two free parameters:
the total normalization and the parameter $g$ characterizing
the level of sea partons in the $\pi$ meson.
The fit gave the following result:

\beq
\label{jpres}
g  \ = \ 1.35 \, \pm \, 0.09\,,
\end{equation}\nsz

\noindent with $\chi^2/NDF\,\approx\,0.9$.
It should be kept in mind, that the parameters
$\alpha$ and $\beta$ have been fixed, $\alpha\,=\,0.5$,
$\beta\,=\,1$.
Results are shown in Fig.\ref{psifit_1}(a-d).

To recalculate the predictions of our model to $NA3$ data on
$pA\,\rightarrow\,J/\psi$ we have to put $N\,=\,3$ in Eq.(\ref{jpsec})
and to change the normalization $\sigma_{tot}^{exp}$, found for the $\pi$
data, by the $W_{p\pi}\,\sigma_{tot}^{exp}$, where
$W_{p\pi}\,=\,\sigma_{tot} (pp\rightarrow J/\psi)/
\sigma_{tot} (\pi p\rightarrow J/\psi)$.
To estimate the ratio $W_{p\pi}$, we neglect any dynamical effects and use
the combinatorial consideration based on the following simple assumptions.
We consider fast quarks only, i.e., the valence and $c$ quarks
of the beam particle. Each massless quark has two spin states and
three colour states (we neglect masses for fast $c$ quarks).
To form white, spin $1$ final $J/\psi$ meson,
we have to choose $c$ ans $\bar c$
quarks with parallel spins in colourless state. It is clear that
the probability of random choice of the $c\bar c$ pair with
necessary quantum numbers is being given by the binomial coefficients
$C_2^n$, where $n$ is the total number of states in the beam particle.
The statistical weight of the final state is the same for both beam
particles and cancels in the ratio.
In $\pi$--mesons we have four fast quarks ($u\bar d
c\bar c$) and, consequently, $n_{\pi}\,=\,3\,\times\,2\,\times\,4=24$ states.
In the proton we have five fast quarks ($uudc\bar c$) and
$n_p\,=\,3\,\times\,2\times\,5=30$ states.
Thus, $W_{p\pi}\,=\,C_2^{24}/C_2^{30}\,=\,92/145$ in good agreement with data.

In Fig.\ref{psifit_1}(e)
we plotted the resulting $d\sigma/dx_F(\JP)$ for $pp$ interactions.
As one can see from Fig.\ref{psifit_1}(e), the model gives
a satisfactory description also for \JP ~production in $pp$ collisions.
This allows us to use the same parameter $g$ for the analysis
of {\em EMC} data on charm production in muon--proton scattering.

Let us remark on some irregularities, seen in
\JP ~distributions in Figs.\ref{psifit_1}(a-c) around the
$x_F\,\approx\,0.8-0.9$. E.g., in Fig.\ref{psifit_1}(c) the point
at $x_f\,\approx\,0.85$ is about one order of magnitude higher, than
the theoretical curve.
If it is not a statistical fluctuation in data, this discrepancy
can be easily understood within the framework of the model \cite{Boreskov93},
which predicts that at very large $A$ and $x_F\,\rightarrow\,1$
the production cross section can have an $A^{1/3}$ dependence, if there is
a final state interaction of the \JP ~particle with nuclear matter.
In this case we would obtain an additional factor
$195^{1/3}\,\approx\,5.8$ in good agreement with Fig.\ref{psifit_1}(c).

In Fig.\ref{psifit_1}(f) we also present the distributions $x\,q(x)$
for charmed and valence quarks in the model and the distribution
of the valence $u$ quark from the $MRS\,(G)$ parameterization.

\section{Charm Electro-Production} 
\label{emcfit}

\bigskip
\subsection{IC Structure Function and Sub--Leading Corrections}

The cross section for charm production in deep inelastic
muon--proton scattering is:

\beq
\label{dissect}
\frac{d\sigma}{dxdQ^2} \ \approx \ \frac{2\pi\,\alpha^2\,[1+(1-y)^2]}{x\,Q^4}\,
F_2^c(x,Q^2)\,,
\end{equation}\nsz

\noindent where, \large{$x\,=\,\frac{Q^2}{2\,(Pq)}$} is the Bjorken variable
and \large{$y\,=\,\frac{Q^2}{sx}$} is the fraction of the muon momentum
carried by the virtual photon (we neglected the contribution
from the longitudinal structure function $F_L$).

The charm structure function of the proton in the present
approach can be represented as a sum of two terms:

\beq
\label{sumsf}
F_2^c(x,Q^2)\ =\ F_2^{(PGF)}(x,Q^2)\ +\ N_c\,F_2^{(IC)}(x,Q^2)\,.
\end{equation}\nsz

\begin{figure}[htb] 
\centerline{\hbox{
\psfig{figure=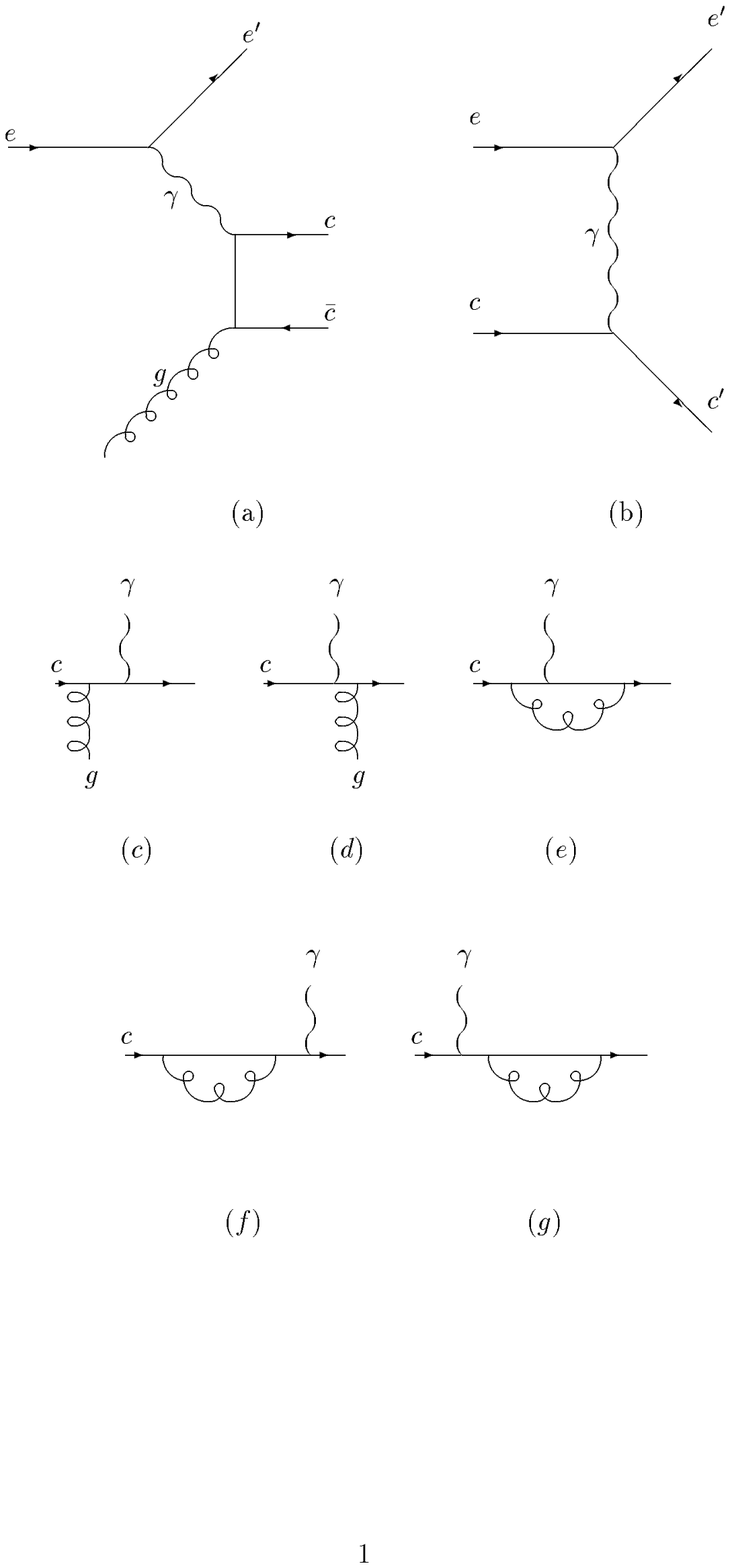,bbllx=5.0cm,bblly=16.5cm,%
bburx=16.0cm,bbury=25.0cm,clip=t,height=6.0cm}
}}
\caption{\label{pgfic}
Diagrams for photon--gluon fusion (a) and the scattering
on the intrinsic charmed quark (b).}
\end{figure}

The first term describes the conventional
photon--gluon fusion mechanism $\mu g\,\rightarrow\,\mu\,c\bar c$
(Fig.\ref{pgfic}a) and the second term represents the direct
scattering of the muon on the constituent charmed quark of the proton,
$\mu\,+\,c\,\rightarrow\,\mu\,+\,c$ (Fig.\ref{pgfic}b). $N_c$ is an unknown
normalization constant to be found from the comparison with
the experimental data.

Within the framework of the naive parton model
the $IC$ structure function is connected with the
distribution $c(x,Q^2)$ of the charmed quark in the proton as:

\beq
\label{f2strf}
F_2^c(x,Q^2)\ =\ 2\,e_c^2\,x\,c(x,Q^2),
\end{equation}\nsz

\noindent $c(x,Q^2)$ is the momentum distribution of the $c$ quark
and $e_c\,=\,\frac{2}{3}$ is the electric charge of the $c$ quark.

\bigskip
The PGF charm structure function is \cite {Gluck87}:

\beq
\label{f2charm}
F_2^{c(PGF)}(x,Q^2)\ =\ \int_{\sqrt{1+4\lambda}\,x}^1
\,\frac{d\xi}{\xi}\,G(\xi,Q^2)\,f_2(\frac{x}{\xi},Q^2),
\end{equation}\nsz

\noindent where,

\beq
\label{f2gluck}
\begin{array}{lll}
f_2(z,Q^2) & = & \frac{\alpha_s(\hat s)}{\pi}\,e_c^2\,{\pi}\,z\,
\biggl \{ V_c\,\left [ -\frac{1}{2}\,+\,2z(1-z)(2-\lambda)\right ] \\
& + & \left [ 1-2z(1-z)\,+\,4\lambda z (1-3z)\,-\,8\lambda^2z^2 \right ]
\ln {\frac{1+V_c}{1-V_c}} \biggr \}.\\
\end{array}
\end{equation}\nsz

In the above expressions
$\hat s\,=\,Q^2\,(1-z)/z$,
$V_c(\hat s)\,=\,\sqrt{1-\,4\,m_c^2/\hat s}$
is the velocity of $c$ quark in the $(\gamma g)$ CM system.

\begin{figure}[htb] 

\centerline{\psfig
{figure=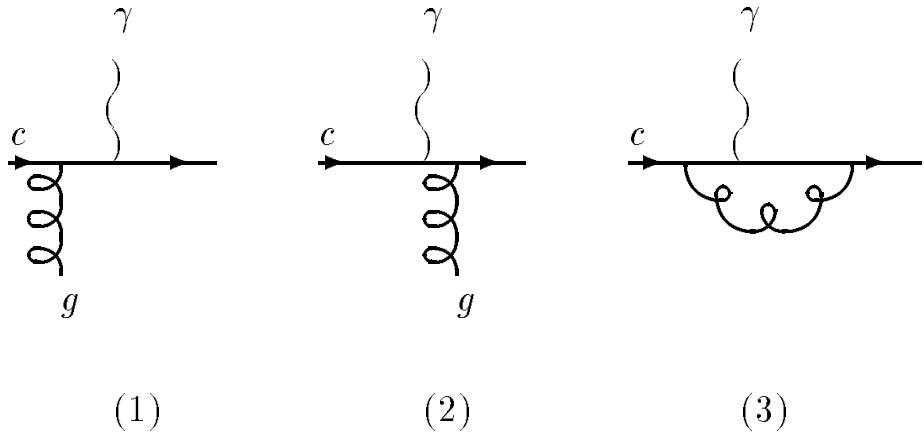,bbllx=4.0cm,bblly=20.4cm,%
bburx=14.0cm,bbury=25.0cm,clip=t,height=6.0cm}
}

\caption{\label{corrs}
$O(\alpha_s)$ corrections to the intrinsic charm structure function:
(1), (2) --- gluon brems\-strah\-lung and (3) --- virtual gluon
corrections.}
\end{figure}

To compare the model with the experimental data
we need to take into account the dependence of the $c(x)$ on
the momentum transfer $Q^2$.
This dependence comes from two sources \cite{Hoffmann83}.
The first one results from the
non--zero masses of the proton and the $c$ quark.
The second source are the first--order QCD radiative corrections
(Fig.\ref{corrs}).

To take into account effects of the non--zero masses
we replaced the variable $x$ by the variable $\zeta$
\cite{Hoffmann83,Barbi76}:

\beq
\label{zeta}
x\ \rightarrow\ \zeta(x)\ =\
\frac{\sqrt{1+4\lambda}+1}{1+\sqrt{1+4\rho x^2}}\,x
\end{equation}\nsz

\npar and the distribution $c(x)$ by the function

\beq
\label{zetascal}
c(x)\ \rightarrow\ c(\zeta,\zh)\ =\
c\left (\frac{\zeta}{\zh}\right ),\ \ \ 0\ \le\ \zeta\ \le\ \zh
\end{equation}\nsz

The parameters $\rho,\,\hat x,\,\lambda,\,\zh$
are then given by:

\beq
\label{params}
\begin{array}{ll}
\rho\ =\ \frac{m_p^2}{Q^2}; & \lambda\ =\ \frac{m_c^2}{Q^2}; \\
\\
x_{max}\ =\ \frac{1}{1+4\lambda -\rho}; & \zh\ =\ \zeta(x_{max}). \\
\end{array}
\end{equation}\nsz


The first--order correction to the structure function
$F_2^{c(0)}(x,Q^2)$ can be
represented as a convolution of the $c$ quark distribution $c(\zeta,\zh)$
with the radiative corrections. As a result the
IC structure function with the radiative corrections has the form
\cite{Hoffmann83}:

\beq
\label{fullstf}
F_2^c (x,Q^2)\ =\ 2\,e_c^2\,\zeta\,c(\zeta,\zh)
+\ 2\,e_c^2\,\zeta\,\int_{\zeta}^{\zh}
\frac{dy}{y}\,c(y,\zh)\sigma_2^{(1)}\left (\frac{\zeta}{y},\lambda\right )
\end{equation}\nsz

The expression for the first--order radiative corrections
$\sigma_2^{(1)} (z,\lambda)$ is given in the Appendix.

In this paper we don't consider the QCD evolution of the constituent
charmed quark distributions and assume that the QCD radiative corrections
to the matrix element of the virtual photon absorption give the correct
description of the $\alpha_s^{(1)}$ effects. At large $Q^2$ this
point needs a more careful study. There is also
the problem of a proper description of the QCD evolution of the heavy
quark distributions at not too large momentum transfers, where
the mass of the heavy partons cannot be neglected.

\subsection{Comparison with {\em EMC} data}

\bigskip
In order to evaluate the contribution to the charm part $F_2^c$
of the proton structure function $F_2$
from the scattering on constituent charm quarks
(also called intrinsic charm quarks) we used the data on
charm production obtained by the {\em EMC} collaboration \cite{EMC82}
in $\mu p$ collisions at $E_{\mu}\,=\,200\ GeV$.
The {\em EMC} collaboration presented data on
\hbox{$\sigma(\gamma^* p\ \rightarrow\ c\bar c\,+\,X)$}.
Thus, we need to extract $F_2^c(x,Q^2)$ from the data
taking into account the difference in definitions of the
virtual photons  flux, being used in Eq.(\ref{dissect})
and by {\em EMC}  collaboration \cite{Hand63} as well as
the finite size of the experimental bins in $(x,Q^2)$ plane.

According to the Equivalent Photon Approximation (EPA)
\cite{Kessler75} the cross section for muon--proton
scattering can be represented as:

\beq
\label{epacrs}
d\sigma(\mu p\,\rightarrow\,c\bar cX)\ =
\ \sigma(\gamma^* p\,\rightarrow\,c\bar cX)\,dn_{\gamma}\,,
\end{equation}\nsz

\noindent where $dn_{\gamma}$ is the differential flux of
equivalent photons. The definition of the equivalent photon
flux is, to some extent, arbitrary. The conventional form,
used in Eq.(\ref{dissect}) for $Q^2/E^2\,\ll\,1$, is of the form:

\beq
\label{gflux}
dn_{\gamma}\ =\ \frac{\alpha}{2\pi}\,
\left [ 1+(1-y)^2\right ]\,\frac{dx}{x}\,\frac{dQ^2}{Q^2}.
\end{equation}\nsz

The {\em EMC} Collaboration used a slightly different definition for the
photon flux \cite{Hand63}, which includes an additional factor
$(1-x)$ in the right--hand--side of Eq.(\ref{gflux}). Taking into account
this factor and approximating the differential flux $dn_{\gamma}/dxdQ^2$
by $\Delta n/\Delta x\Delta Q^2$
(where $(\Delta \nu,\Delta Q^2)$ is the experimental bin),
we can relate the structure function to
the experimentally measured $\gamma p$ cross section,
$\sigma_{\gamma p}^{exp}$, for charm production:

\beq
\label{exptrf}
F_2^c(x,Q^2)\ =\ \frac{\Delta n_{\gamma}}{\Delta \nu\,\Delta Q^2}\
\frac{\nu\,Q^4}{4\pi\alpha\left [1+(1-y)^2\right ]}\,.
\sigma_{\gamma}^{exp}(\nu ,Q^2)
\end{equation}\nsz

The value of $\Delta n_{\gamma}$ is found by integrating the expression
for the photon flux over each experimental bin $(\Delta \nu,\Delta
Q^2)$.
Defining $\Delta\nu\,=\,\nu_2-\nu_1$ and $\Delta Q^2\,=\,Q_1^2-Q_2^2$,
we obtain:

\beq
\label{dltflx}
\begin{array}{lll}
\Delta n_{\gamma}(\nu ,Q^2) & = & \frac{\alpha}{2\pi}\,
\left\{
 \ln \left (\frac{Q_2^2}{Q_1^2}\right )
\left [ 2\ln\frac{\nu_2}{\nu_1}-\frac{2}{E}(\nu_2-\nu_1)+
\frac{\nu_2^2-\nu_1^2}{2E^2}\right ] \right .\\
\\
 & &
\left .
-\,\frac{Q_2^2-Q_1^2}{s}\,\left [ 2E(\frac{1}{\nu_1}-\frac{1}{\nu_2})
-2\ln \frac{\nu_2}{\nu_1}\,+\,\frac{\nu_2-\nu_1}{E}\right ]
\right\}\,.
\end{array}
\end{equation}\nsz

\bigskip
Mass corrections for IC structure functions have been used
in the form of $\zeta$--scaling (\ref{zetascal}).
The radiative corrections (see Appendix)
have also been taken into account.
The strong coupling constant $\alpha_s(Q^2)$
and the $\Lambda_{QCD}$ has been chosen according to the PDFLIB
parameterization \cite{PDFlib}.

To perform a fit to {\em EMC} data we used expression (\ref{sumsf})
for the charmed structure function with two free parameters:
$N_c$ being the normalization of $F_2^{(IC)}$
and $m_c$ the mass of charmed quark, entering the PGF structure
function and the radiative and mass corrections.
For the gluon distribution we have chosen the {\em MRS(G)}
parameterization, which is the default one for PDFLIB~7.09 \cite{PDFlib}.

\begin{figure}[htb] 
\centerline{\hbox{
\psfig{figure=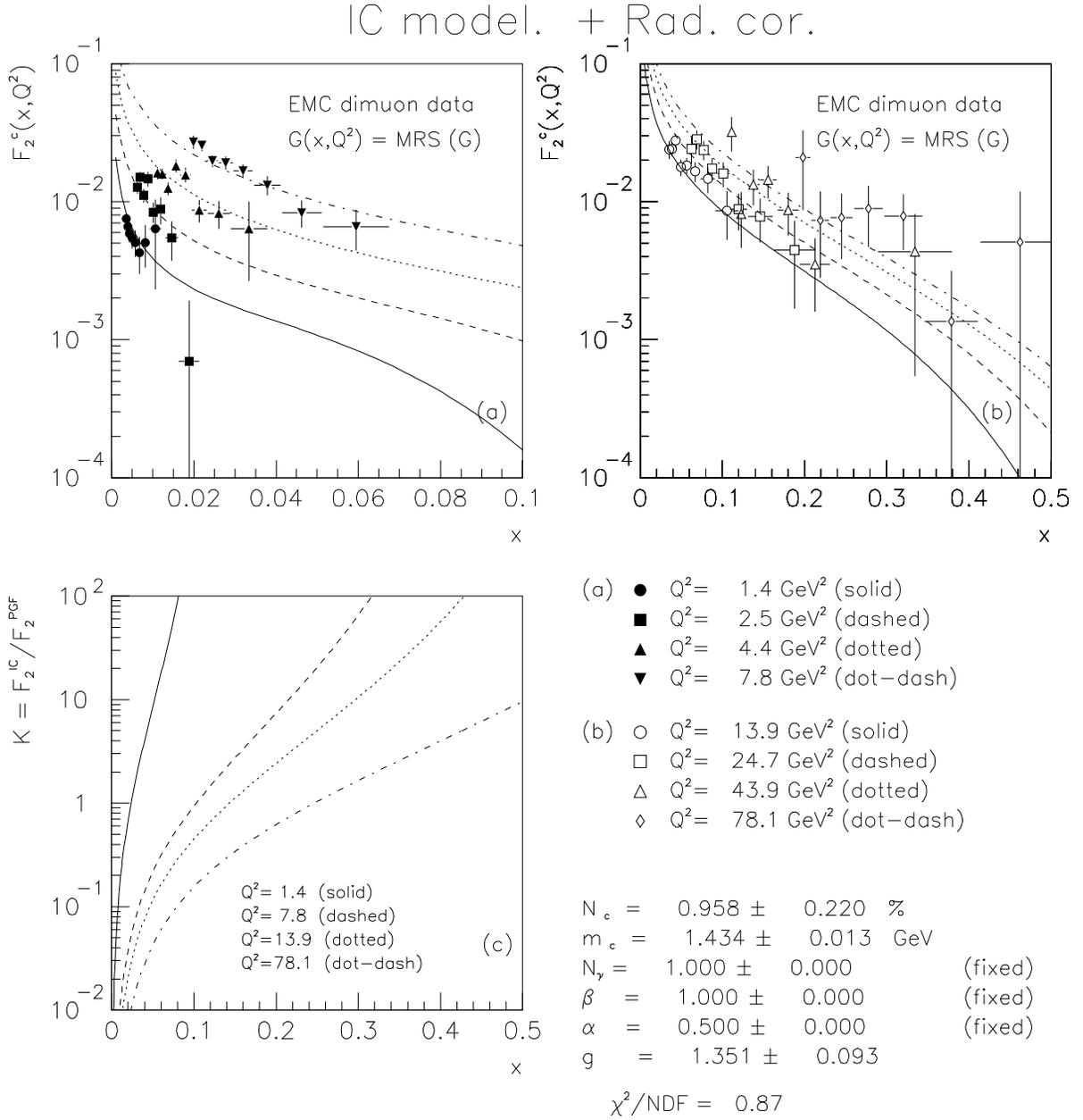,bbllx=0.7cm,bblly=5.2cm,%
bburx=19.0cm,bbury=23.0cm,clip=t,height=16.0cm}
}}
\caption{\label{emcres1}
Results of the fit of the sum PGF + IC to the {\em EMC} data
(a,b) and the ratio $K\ =\ F_2^{IC}/F_2^{PGF}$ of the contributions
from intrinsic charm and photon gluon fusion (c).}
\end{figure}

\bigskip
The results of our fit to the {\em EMC} data are shown
in Fig.\ref{emcres1}(a,b).
The fit yields for the admixture from scattering on
the constituent charmed quark

\bmath
N_c\ =\ (0.9\,\pm\,0.2)\,\%
\emath

\noindent and a value of the mass of

\bmath
m_c\ =\ (1.43\,\pm\,0.01)\ GeV
\emath

\noindent for the charmed quark mass.
We have used also other PDFLIB parameterizations and, within errors,
obtained similar values.

Figure \ref{emcres1}(c) presents the ratio $F_2^{IC}/F_2^{PGF}$
versus $x$ for some values of $Q^2$. We see that at large $x$
($x\,>\,0.1$) an $1\%$ IC component dominates the charm production for
$Q^2\,\lesssim\,(10-12)\ GeV^2$.

\section{Conclusion} 
\label{concl}

\bigskip
In this paper we modified and generalized the statistical approach
to the Fock state hadron structure with heavy quarks,
using the framework of the non--covariant perturbation theory.
We obtained scaling expressions for heavy and light parton
distributions in the infinite momentum frame.
We calculated the charmed structure function of the proton
$F_2^{(c)}(x,Q^2)$ taking into account the QCD radiative corrections
of order $\alpha_s^{(1)}$ as well as the mass corrections to the
structure function caused by the non--zero values of
the $c$ quark and the proton masses.
We used the experimental data on $\pi A\,\rightarrow\,\JP X$
and $\mu p\,\rightarrow\,\mu c\bar c X$
to evaluate the parameters of the model and presented the relative
contributions of the photon--gluon fusion mechanism and the direct
scattering on the constituent charmed quark (intrinsic charm) to the
charmed structure function of the proton.

We found that in $\pi A$ collisions the so called "diffractive"
component of \JP ~can be well described by the coalescing of
$c$ and $\bar c$ constituent quarks. This success supports,
in our view, Gribov's space--time picture of the
hadron interaction with nuclei as well as the presence of
long--living heavy quark fluctuations in the hadron ("intrinsic charm").
The results show also that the longitudinal distribution
of the constituent heavy quarks is harder than for
light sea partons and has a shape like that of the valence
quark distributions.

From the comparison with the charm muon--production at
$E_{\mu}\,=\,200\ GeV$ we estimated the contribution from scattering
on constituent charmed quarks to the total charm production.
This contribution is about $1\%$ and is expected to grow with
the beam energy. At large values of the Bjorken variable $x\,\gtrsim\,0.1$
the scattering on the constituent charmed quark dominates the forward
charm production in deep inelastic lepton--proton collisions.

The considered approach taken for the excitation of Fock states,
containing heavy quarks, can also be applied
to states with beauty quarks in the hadron as well as
to the hadronic component of the virtual photon ("resolved photon").

Finally we want to underline that the $ep$ collider HERA
is well suited to investigate heavy flavour
production mechanisms over a wide kinematic region of $(x,Q^2)$
which is inaccessible at other existing facilities.

\bigskip 
\vbox{
\centerline{\bf{Acknowledgment}}

I want to thank Prof.~G.Wolf very much for
illuminating discussions and exceptionally useful remarks.
I would like to express my gratitude to DESY for partial support
during this work.

I want also to thank Dr.~B.Harris for mutual cross--check of the
codes for calculation of the radiative corrections.
}

\vfill\eject



\vfill\eject
\centerline{\large {\bf APPENDIX}}

\bigskip

The expression for the first--order radiative corrections
$\sigma_2^{(1)} (z,\lambda)$ to the charmed structure function is:

\begin{equation}
\label{radcor}
\begin{array}{lll}
\sigma_2^{(1)} (z,\lambda)\ & =\ &  \frac{2\alpha_s}{3\pi}\delta(1-z)\,
\biggl \{ 4\ln\lambda-2+\sqrt{1+4\lambda}\,L\\
\\
&+&\frac{(1+2\lambda)}{\sqrt{1+4\lambda}} \left [3L^2+4L+4Li_2(\frac{-d}{a})
+2L\ln \lambda -4L\ln (1+4\lambda)+2Li_2(\frac{d^2}{a^2})\right] \biggr \}\,\\
\\
&+& \frac{\alpha_s}{3\pi}\frac{1}{(1+4\lambda z^2)^2}\,\biggl \{\frac{1}
{[1-(1-\lambda)z]^2}\,\bigl [ (1-z)(1-2z-6z^2+8z^4)\, \\
\\
&+& 6\lambda z(1-z)(3-15z-2z^2+8z^3)\,+\,4\lambda^2z^2(8-77z+65z^2-2z^3)\,+ \\
\\
&+& 16\lambda^3z^3(1-21z+12z^2)-128\lambda^4z^5 \bigr ]\\
\\
&-& \frac{2\hat L}{\sqrt{1+4\lambda z^2}} \bigl [ (1+z)(1+2z^2)-2\lambda z
(2-11z-11z^2)-8\lambda ^2z^2(1-9z)\bigr ] \\
\\
&-& \frac{8z^4(1+\lambda )^2}{(1-z)_+}\,-\,\frac{4z^4(1+2\lambda)
(1+4\lambda)^2\hat L}{\sqrt{1+4\lambda z^2}(1-z)_+} \biggr \},\\
\end{array}
\end{equation}

\noindent with the definitions:

\begin{equation}
\begin{array}{lll}
\hat L & = & \ln \frac{4\lambda z [1-(1-\lambda)z]}{(1+2\lambda z +
\sqrt{1+4\lambda z^2})^2};\\
\\
a & = &\frac{\sqrt{1+4\lambda}+1}{2};\\
\\
d & = & a-1;\\
\\
L & = & \ln\frac{a}{d};\\
\\
Li_2 & = & -\int_0^x{dz\,\frac{\ln (1-z)}{z}};\\
\end{array}
\end{equation}

\vfill\eject

\vfill\eject


\vfill\eject

\vfill\eject

\end{document}